\documentclass[
superscriptaddress,
 amsmath,amssymb,
 aps
]{revtex4-1}

\usepackage{upgreek}
\usepackage{lineno}
\usepackage{graphicx}
\usepackage{dcolumn}
\usepackage{bm}
\usepackage[scr]{rsfso}

\usepackage{subfigure}
\usepackage{orcidlink}
\usepackage{xspace}
\usepackage{ifthen}
\usepackage{xcolor}
\newboolean{uprightparticles}
\setboolean{uprightparticles}{false}

\def\SND {\mbox{SND@LHC}\xspace}

\def\ux85 {\mbox{UX85}\xspace}

\ifthenelse{\boolean{uprightparticles}}%
{

 \def\Pmu         {\ensuremath{\upmu}\xspace}
 \def\Pnu         {\ensuremath{\upnu}\xspace}

 \def\PDelta      {\ensuremath{\Delta}\xspace}
 \def\PXi      {\ensuremath{\Xi}\xspace}
 \def\PLambda      {\ensuremath{\Lambda}\xspace}
 \def\PSigma      {\ensuremath{\Sigma}\xspace}
 \def\POmega      {\ensuremath{\Omega}\xspace}
 \def\PUpsilon      {\ensuremath{\Upsilon}\xspace}

 \def\PB      {\ensuremath{\mathrm{B}}\xspace}
 
 \def\PD      {\ensuremath{\mathrm{D}}\xspace}

 \def\PK      {\ensuremath{\mathrm{K}}\xspace}

 \def\Pi      {\ensuremath{\mathrm{i}}\xspace}

}
{

 \def\Pmu         {\ensuremath{\mu}\xspace}
 \def\Pnu         {\ensuremath{\nu}\xspace}

 \mathchardef\PDelta="7101
 \mathchardef\PXi="7104
 \mathchardef\PLambda="7103
 \mathchardef\PSigma="7106
 \mathchardef\POmega="710A
 \mathchardef\PUpsilon="7107
 
 \def\PB      {\ensuremath{B}\xspace}
 
 \def\PD      {\ensuremath{D}\xspace}

 \def\PK      {\ensuremath{K}\xspace}

 \def\Pi      {\ensuremath{i}\xspace}

}

\def\mup        {\ensuremath{\Pmu^+}\xspace}
\def\mun        {\ensuremath{\Pmu^-}\xspace} 

\def\neu        {\ensuremath{\Pnu}\xspace}

\def\neum       {\ensuremath{\neu_\mu}\xspace}

\def\kaon  {\ensuremath{\PK}\xspace}
  \def\Kbar  {\kern 0.2em\overline{\kern -0.2em \PK}{}\xspace}

\def\Kz    {\ensuremath{\kaon^0}\xspace}
\def\Kzb   {\ensuremath{\Kbar^0}\xspace}
\def\KzKzb {\ensuremath{\Kz \kern -0.16em \Kzb}\xspace}
\def\Kp    {\ensuremath{\kaon^+}\xspace}
\def\Km    {\ensuremath{\kaon^-}\xspace}

\def\KpKm  {\ensuremath{\Kp \kern -0.16em \Km}\xspace}

\def\KL    {\ensuremath{\kaon^0_{\rm\scriptscriptstyle L}}\xspace}

  \def\Dbar    {\kern 0.2em\overline{\kern -0.2em \PD}{}\xspace}
\def\D       {\ensuremath{\PD}\xspace}

\def\Dz      {\ensuremath{\D^0}\xspace}
\def\Dzb     {\ensuremath{\Dbar^0}\xspace}
\def\DzDzb   {\ensuremath{\Dz {\kern -0.16em \Dzb}}\xspace}
\def\Dp      {\ensuremath{\D^+}\xspace}
\def\Dm      {\ensuremath{\D^-}\xspace}

\def\DpDm    {\ensuremath{\Dp {\kern -0.16em \Dm}}\xspace}

\def\Bbar    {\ensuremath{\kern 0.18em\overline{\kern -0.18em \PB}{}}\xspace}

  \def\Y#1S{\ensuremath{\PUpsilon{(#1S)}}\xspace}

\def\Lbar {\ensuremath{\kern 0.1em\overline{\kern -0.1em\PLambda}}\xspace}

\def\AT#1     {\ensuremath{A_{\mathrm{T}}^{#1}}\xspace}           

\def\C#1      {\ensuremath{\mathcal{C}_{#1}}\xspace}                       
\def\Cp#1     {\ensuremath{\mathcal{C}_{#1}^{'}}\xspace}                    
\def\Ceff#1   {\ensuremath{\mathcal{C}_{#1}^{\mathrm{(eff)}}}\xspace}        
\def\Cpeff#1  {\ensuremath{\mathcal{C}_{#1}^{'\mathrm{(eff)}}}\xspace}       
\def\Ope#1    {\ensuremath{\mathcal{O}_{#1}}\xspace}                       
\def\Opep#1   {\ensuremath{\mathcal{O}_{#1}^{'}}\xspace}                    

\newcommand{\tev}{\ensuremath{\mathrm{\,Te\kern -0.1em V}}\xspace}
\newcommand{\gev}{\ensuremath{\mathrm{\,Ge\kern -0.1em V}}\xspace}
\newcommand{\mev}{\ensuremath{\mathrm{\,Me\kern -0.1em V}}\xspace}
\newcommand{\kev}{\ensuremath{\mathrm{\,ke\kern -0.1em V}}\xspace}
\newcommand{\ev}{\ensuremath{\mathrm{\,e\kern -0.1em V}}\xspace}
\newcommand{\gevc}{\ensuremath{{\mathrm{\,Ge\kern -0.1em V\!/}c}}\xspace}
\newcommand{\mevc}{\ensuremath{{\mathrm{\,Me\kern -0.1em V\!/}c}}\xspace}
\newcommand{\gevcc}{\ensuremath{{\mathrm{\,Ge\kern -0.1em V\!/}c^2}}\xspace}
\newcommand{\gevgevcccc}{\ensuremath{{\mathrm{\,Ge\kern -0.1em V^2\!/}c^4}}\xspace}
\newcommand{\mevcc}{\ensuremath{{\mathrm{\,Me\kern -0.1em V\!/}c^2}}\xspace}

\def\gsim{{~\raise.15em\hbox{$>$}\kern-.85em
          \lower.35em\hbox{$\sim$}~}\xspace}
\def\lsim{{~\raise.15em\hbox{$<$}\kern-.85em
          \lower.35em\hbox{$\sim$}~}\xspace}

\def\tell1  {TELL1\xspace}
\def\ukl1   {UKL1\xspace}

\begin{document}

\title{Observation of collider muon neutrinos  with the \SND experiment}


\author{R.~Albanese~\orcidlink{0000-0003-4586-8068}}
\affiliation{Sezione INFN di Napoli, Napoli, Italy}
\affiliation{Universit\`{a} di Napoli ``Federico II'', Napoli, Italy}

\author{A.~Alexandrov~\orcidlink{0000-0002-1813-1485}}
\affiliation{Sezione INFN di Napoli, Napoli, Italy}

\author{F.~Alicante~\orcidlink{0009-0003-3240-830X}}
\affiliation{Sezione INFN di Napoli, Napoli, Italy}
\affiliation{Universit\`{a} di Napoli ``Federico II'', Napoli, Italy}

\author{A.~Anokhina~\orcidlink{0000-0002-4654-4535}}
\affiliation{Affiliated with an institute covered by a cooperation agreement with CERN}

\author{T.~Asada~\orcidlink{0000-0002-2482-8289}}
\affiliation{Sezione INFN di Napoli, Napoli, Italy}
\affiliation{Universit\`{a} di Napoli ``Federico II'', Napoli, Italy}

\author{C.~Battilana~\orcidlink{0000-0002-3753-3068}}
\affiliation{Sezione INFN di Bologna, Bologna, Italy}
\affiliation{Universit\`{a} di Bologna, Bologna, Italy}

\author{A.~Bay~\orcidlink{0000-0002-4862-9399}}
\affiliation{Institute of Physics, \'{E}cole Polytechnique F\'{e}d\'{e}rale de Lausanne (EPFL), Lausanne, Switzerland}

\author{C.~Betancourt~\orcidlink{0000-0001-9886-7427}}
\affiliation{Physik-Institut, Universit\"{a}t Z\"{u}rich, Z\"{u}rich, Switzerland}

\author{A.~Blanco~Castro~\orcidlink{0000-0001-9827-8294}}
\affiliation{Laboratory of Instrumentation and Experimental Particle Physics (LIP), Lisbon, Portugal}

\author{M.~Bogomilov~\orcidlink{0000-0001-7738-2041}}
\affiliation{Faculty of Physics, Sofia University, Sofia, Bulgaria}

% ...

\collaboration{\SND~Collaboration}

\date{\today}

\begin{abstract}
We report the direct observation of muon neutrino interactions with the \SND detector at the Large Hadron Collider. A data set of proton-proton collisions at $\sqrt{s} = 13.6\,$TeV collected by \SND in 2022 is used, corresponding to an integrated luminosity of 36.8$\,\rm{fb}^{-1}$. The search is based on information from the active electronic components of the \SND detector, which covers the pseudo-rapidity region of $7.2 < \eta < 8.4$, inaccessible to the other experiments at the collider. Muon neutrino candidates are identified through their charged-current interaction topology, with a track propagating through the entire length of the muon detector. After selection cuts, 8 \neum interaction candidate events remain with an estimated background of 0.076 events, yielding a significance of seven standard deviations for the observed \neum signal.
\begin{center}
\begin{normalsize}
\textcopyright 2023 CERN for the benefit of the \SND Collaboration. Reproduction of this article or parts of it is allowed as
specified in the CC-BY-4.0 license
\end{normalsize}
\end{center}
\end{abstract}

\maketitle


\label{sec:intro}
\vspace{0.3cm}
\twocolumngrid

{\it Introduction -} 
The use of the Large Hadron Collider (LHC) as a neutrino factory was first envisaged about 30 years ago~\cite{DeRujula:1984pg,DeRujula:1992sn,Vannucci:253670}
in particular for the then undiscovered $\nu_{\tau}$~\cite{Jarlskog:215298}.
Those studies suggest a detector intercepting the very forward flux (${\eta > 7}$) of neutrinos (about 5\% have $\tau$ flavour) from $b$ and $c$ decays~\cite{Buontempo:2018gta}.
The physics potential of a detector to study neutrinos was underlined in Ref.~\cite{Beni:2019gxv}. The role of an off-axis setup, which enhances the neutrino flux from charmed particle decays, was emphasised in Ref.~\cite{Beni:2020yfy}. 

Proton-proton ({\it pp}) collisions at a center-of-mass energy of 13.6~\tev during LHC Run 3, with an expected integrated luminosity of $250~\rm{fb}^{-1}$, will produce a high-intensity beam yielding O($10^{12}$) neutrinos in the far forward direction with energies up to a few~\tev\cite{Ahdida:2750060}. 

Neutrinos allow precise tests of the Standard Model (SM)~\cite{brock,conrad,formaggio,delellis} and are a probe for 
new physics~\cite{marfatia,arguelles}. Measurements of the neutrino cross section in the last decades were mainly performed at low energies. The region between 350 \gev and 10 \tev is currently unexplored~\cite{bustamante}. 

\SND~\cite{snddet} was designed to perform measurements with high-energy neutrinos (100~\gev to a few~\tev) 
produced at the LHC in the pseudo-rapidity region $7.2<\eta<8.4$. It is a compact, standalone experiment located in the TI18 unused LEP transfer tunnel (480~m away of the ATLAS interaction point, IP1\cite{ATLAS:2016fhk}) where it is shielded from collision debris by around 100~m of rock and concrete. The detector is capable of identifying all three neutrino flavours with high efficiency. 

The detector was installed in TI18 in 2021 during the Long Shutdown~2 and has collected data since the beginning of the LHC Run~3 in April 2022. The experiment will run throughout the whole Run~3, during which a total of two thousand high-energy neutrino interactions of all flavours are expected to occur in the detector target.


In this paper, we report the detection of $\nu_\mu$ charged-current (CC) interactions using only data that was taken by the electronic detectors in 2022. 

Recently the observation of neutrino interactions in a complementary pseudo-rapidity region ($\eta>8.8$) has also been reported with the analysis of the 2022 data by the FASER Collaboration~\cite{fasercollaboration2023direct}.

\label{sec:detector}
\vspace{0.3cm}

\begin{figure}[htb]
\centering
\includegraphics[width=1.0\textwidth]{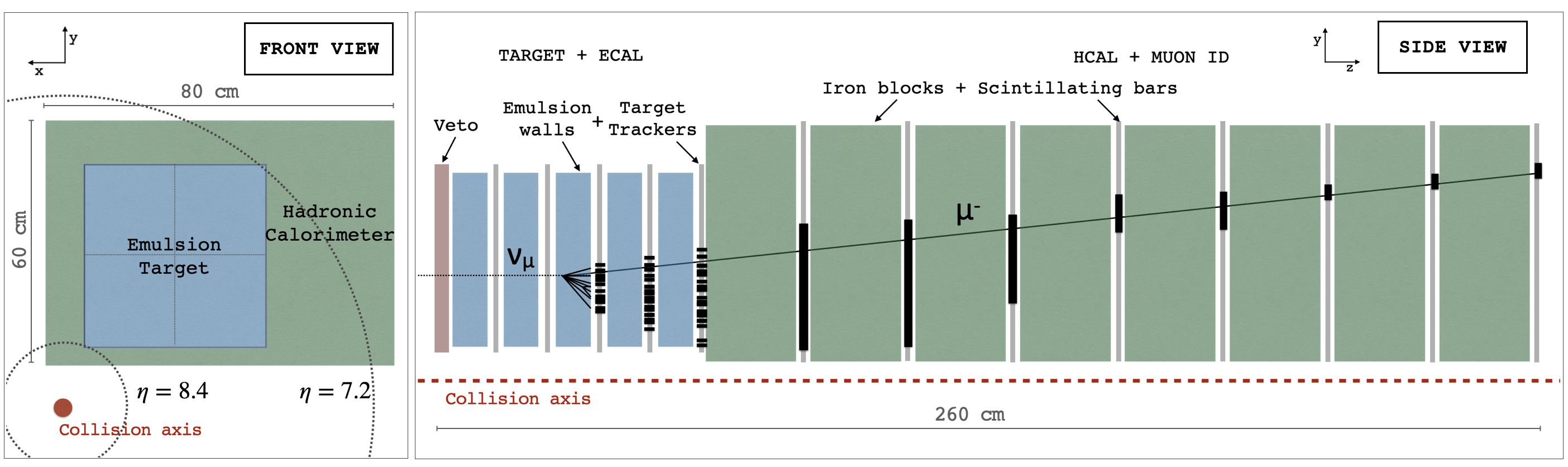}
\caption{Schematic layout of the SND@LHC detector front view (left) and side view (right). The side view includes an illustration of a $\nu_\mu$ charged-current interaction in the target with a hadronic shower sampled in the emulsion target, target trackers, and hadronic calorimeter, and a muon track visible in the muon system.}
\label{fig:detector}
\end{figure}

{\it Detector -} The \SND detector consists of a hybrid system with a $\sim830\,$kg target made of tungsten plates interleaved with nuclear emulsions and electronic trackers, followed by a hadronic calorimeter and a muon system (see Figure~\ref{fig:detector}). The electronic detectors provide the time stamp of the neutrino interaction, preselect the interaction region, tag muons and measure the electromagnetic and hadronic energy, while the emulsion detectors provide excellent vertex reconstruction.

A left-handed coordinate system is used, with $z$ along the nominal collision axis and pointing away from IP1, $x$ pointing towards the center of the LHC, and $y$ vertically aligned and pointing upwards.

The detector consists of three parts: the veto system, the target section, and the hadronic calorimeter and muon system. 

The veto system is located upstream of the target region and comprises two parallel planes, located $4.3\,$cm apart, of scintillating bars read out on both ends by silicon photomultipliers (SiPMs). Each plane consists of seven $1 \times 6 \times 42\,$cm$^3$ stacked bars of EJ-200 scintillator~\cite{ej200}.
The number of photons generated by a minimum-ionising particle crossing $1\,$cm scintillator is of the order of $10^4$.
The bars are wrapped in aluminized BoPET foil to ensure light tightness and therefore isolate them from light in adjacent bars. This system is used to tag muons and other charged particles entering the detector from the IP1 direction.

The target section contains five walls. Each wall consists of four 
units of emulsion cloud chambers (ECC~\cite{emulsions}) and is followed by a scintillating fibre (SciFi~\cite{LHCb_Tracker_TDR}) station for tracking and electromagnetic calorimetry.

\vspace{3mm}
\onecolumngrid

\twocolumngrid

The sub-micrometric spatial resolution of the nuclear emulsions allows for very efficient tracking of all the charged particles produced in high energy neutrino interactions, despite their small angular separation due to the large boost. This also allows for efficient tracking of the tau lepton and its decay vertex which in turn is a key element in the identification of tau leptons and hence the tagging of tau neutrino interactions.

Each ECC unit is a sequence of 60 nuclear emulsion 
films, $19.2 \times 19.2\,$cm$^2$ and approximately 300~$\upmu$m, interleaved with 59 tungsten plates, 1~mm thick. 
Its weight is approximately 41.5~kg, adding up to about 830~kg for the total target mass. 

Each SciFi station consists of two $40 \times 40\,$cm$^2$ planes, alternating $x$ and $y$ views. Each view comprises six densely packed staggered layers of 250$\,\upmu$m diameter polystyrene-based scintillating fibres read out by SiPM arrays. The single particle spatial resolution in one view is of order of $\sim150\,\upmu$m and the time resolution for a particle crossing both $x$ and $y$ views of one plane is about $250\,$ps.

The muon system and hadronic calorimeter consists of two parts: upstream (US), the first five stations,  and downstream (DS), the last three stations (see Figure~\ref{fig:detector}). Each US station consists of 10 stacked horizontal scintillator bars of $82.5 \times 6 \times 1\,$cm$^3$, similar to the  veto detector, resulting in a coarse $y$ view. A DS station consists of two layers of thinner $82.5 \times 1 \times 1\,$cm$^3$ bars arranged in alternating $x$ and $y$ views, allowing for a spatial resolution in each view of less than $1\,$cm. The time resolution for a single DS detector bar is $\sim120\,$ps. The eight scintillator planes are interleaved with $20\,$cm thick iron blocks. In combination with SciFi, the muon system and hadronic calorimeter acts as a coarse sampling calorimeter ($\sim9.5\,\lambda_{\rm{int}}$ in the US detector), providing the energy measurement of hadronic jets. The finer spatial resolution of the DS detector allows for the identification of muon tracks exiting the detector.

All signals exceeding preset thresholds are read out by the front-end electronics and clustered in time to form events. A software noise filter is applied to the events online, resulting in negligible detector deadtime or loss in signal efficiency. Events satisfying certain topological criteria, such as the presence of hits in several detector planes, are read out at a rate of around 5.4 kHz at the highest instantaneous luminosity achieved in 2022 of $2.5\times10^{34}\,$cm$^{-2}$ s$^{-1}$.

\label{sec:data}
\vspace{0.3cm}

{\it Data and simulated samples - } 
In this paper, we analyse the data collected during 2022, with {\it pp} collisions at center of mass energy of 13.6 \tev.
The delivered integrated luminosity during this period, as estimated by the ATLAS Collaboration\cite{ATLAS:2016fhk,ATLAS:2022hro}, was 38.7$\,\rm{fb}^{-1}$, of which 36.8$\,\rm{fb}^{-1}$ were recorded, corresponding to a detector uptime of 95\%. The data set comprises a total of 8.3$\times 10^{9}$ events.

The analysis developed for the first observation of \neum CC interactions from LHC collisions is conducted solely using the data from the electronic detectors, as information from the emulsion target is currently being processed.

In \SND the dominant CC process occurring for $\nu_{\mu}$s is deep inelastic scattering (CCDIS), given the high energy of neutrinos within the detector acceptance~\cite{Ahdida:2750060}.
The signature of these interactions includes an isolated muon track in the muon system, associated with a hadronic shower detected in the SciFi and hadronic calorimeter. In Figure~\ref{fig:detector} the distinctive topology of \neum CCDIS interactions is shown.

Neutrino production in $pp$ collisions at the LHC is simulated with the \textsc{FLUKA} Monte Carlo code~\cite{Fluka2,Fluka3}. \textsc{DPMJET3} (Dual Parton Model, including charm)~\cite{Roesler_2001,DPMJET}  is used for the $pp$ event generation, and \textsc{FLUKA} performs the particle propagation towards the SND@LHC detector with the help of a detailed simulation of LHC accelerator elements~\cite{Boccone:2014hxd}. \textsc{FLUKA} also takes care of simulating the production of neutrinos from decays of long-lived products of the $pp$ collisions and of particles produced in re-interactions with the surrounding material. \textsc{Genie}~\cite{cite:GENIE} is then used to simulate neutrino interactions with the detector material. The propagation of particles through the TI18 tunnel and the \SND detector is simulated with \textsc{Geant4}. A total of around 160 thousand simulated neutrino events and 30 million background events were generated for the analysis described in this publication. 

Given the total mass of the tungsten target during the 2022 run ($\sim800\,$kg), about 157 $\pm$ 37 \neum CCDIS interactions are expected in the full target in the analysed data set, where the range in the expectation is given by the difference between the predictions of the \neum flux at \SND using \textsc{DPMJET3} and \textsc{SIBYLL} obtained in Ref.\cite{Kling:2021gos}.

\vspace{0.3cm}

\begin{figure}[b]
\centering
\includegraphics[width=0.5\textwidth]{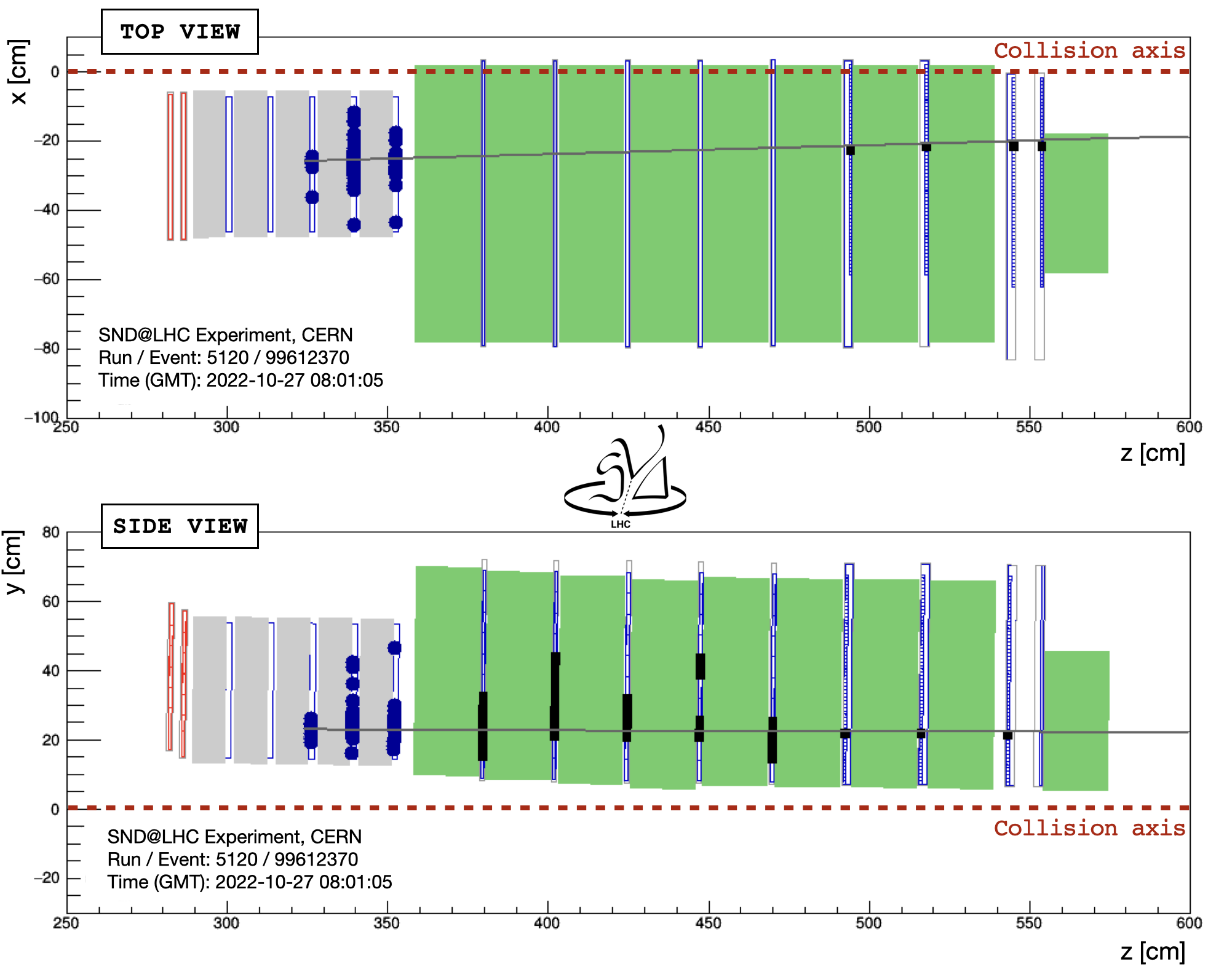}
\caption{Display of a \neum CC candidate event. Hits in the SciFi, and hadronic calorimeter and muon system are shown as blue markers and black bars, respectively, and the line represents the reconstructed muon track.}
\label{fig:candidate}
\end{figure}

{\it Analysis -}
Observing the rare neutrino signal over the prevailing background implies adopting a selection with strong rejection power, designed to yield a clean set of events.

The signal selection proceeds in two steps. 
The first step aims at identifying events happening in a fiducial region of the target, while rejecting backgrounds due to charged particles entering from the front and sides of the detector. Cuts are applied on the hit multiplicity in the veto and SciFi planes to select events that are located in the $3^{\mathrm{rd}}$ or $4^{\mathrm{th}}$ target wall and consistent with a neutral particle interaction. The exclusion of events starting in the two most upstream target walls enhances the rejection power for muon-induced backgrounds, while excluding events starting in the most downstream wall ensures the neutrino-induced showers are sampled by at least two SciFi planes. The average SciFi channel and DS bar number are used to discard events with hits at the edges of detectors' sensitive areas, resulting in a fiducial cross-sectional area in the $xy$ plane of $25 \times 26\,$cm$^{2}$.  The efficiency of  fiducial region cuts on simulated neutrino interactions in the target is $7.5\%$.

The second step selects signal-like signature patterns using a cut-based procedure. \neum CCDIS interactions are associated to a large hadronic activity in the calorimetric system, with a clean outgoing muon track reconstructed in the muon system, and hit time distribution consistent with an event originating from the IP1 direction. The muon track is defined by a set of muon system hits in a straight-line pattern spanning at least three detector planes in both $zx$ and $zy$ views. Events with a large number of hits in the muon system are rejected to ensure cleanly reconstructed tracks.

The achieved reduction factor on the data for the total selection (fiducial and neutrino identification cuts) amounts to $1.0\times 10^{9}$, while the overall efficiency on the  \neum CCDIS Monte Carlo sample is $2.9\%$. 

\begin{figure}[b]
\centering
\includegraphics[width=0.4\textwidth]{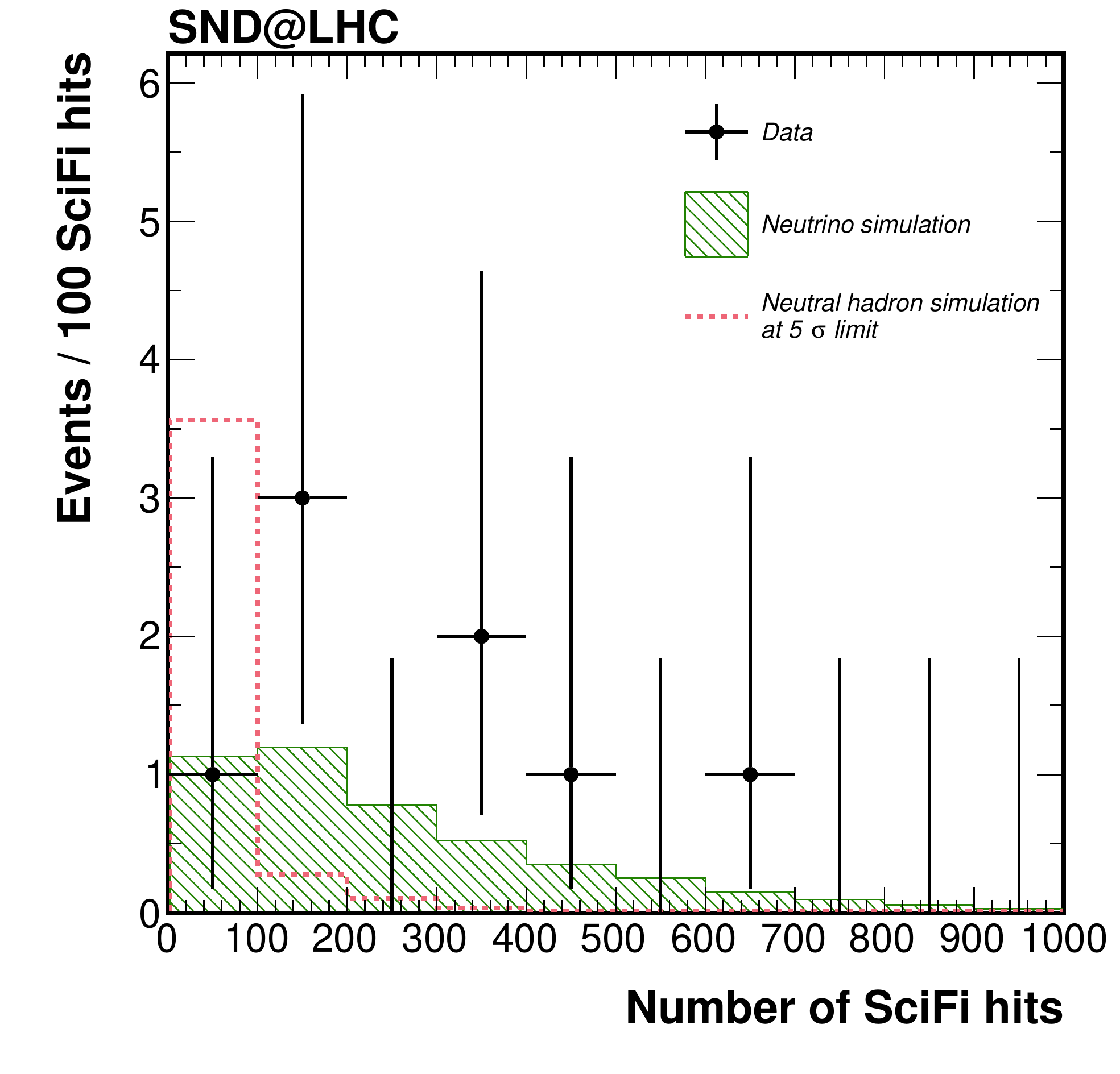}
\caption{Distribution of SciFi hits for candidate events, along with the expectation from the neutrino signal. The dashed line shows the background-only hypothesis scaled up to a deviation from the nominal expectation at a level of five standard deviations. The vertical bars represent 68.3\% confidence intervals of the Poisson means.}
\label{fig:scifi_hits}
\end{figure}

As a result of the full selection, 8 \neum CCDIS candidates are identified, while 4.5 are expected. The contribution of other neutrino flavours and neutral current interactions to the selected sample is less than 1\% of the expected \neum CCDIS yield.
One of the selected candidates is shown in Figure~\ref{fig:candidate}. The distribution of the number of hits in the SciFi detector for the selected events is consistent with the neutrino signal expectation, as shown in Figure~\ref{fig:scifi_hits}.

\vspace{0.3cm}
{\it Background -}
Muons reaching the detector location are the main source of background for the neutrino search. They can either enter in the fiducial volume without being vetoed and generate showers via bremsstrahlung or deep inelastic scattering, or interact in the surrounding material and produce neutral particles that can then mimic neutrino interactions in the target.

The estimate of the penetrating muon background is based on the expected flux in the fiducial volume and on the inefficiency of detector planes used as veto: the veto system and the two most upstream SciFi planes.

The muon flux at the detector location is estimated by the CERN SY-STI team with simulations of proton-proton interactions at IP1 and the transport of the resulting charged pions and kaons along the LHC straight section until their decay using \textsc{FLUKA}~\cite{Fluka2,Fluka3}. The simulation includes both the effects of the accelerator optics and of the material traversed by the particles before reaching the detector. The muons are recorded at a scoring plane, $1.8\times1.8\,\mathrm{m}^2$, located about $70\,$m upstream of \SND, $419\,$m from IP1. The transport of muons from the scoring plane to the detector and their interactions along the way are modelled with a \textsc{Geant4} simulation of \SND and its surroundings.
The \textsc{FLUKA} simulation consists of $50$ million {\it pp} collisions with LHC Run 3 beam conditions and a downward crossing angle of $-160~\rm{\upmu rad}$ on the vertical plane, corresponding to the collider configuration in 2022.

The expected muon flux in the fiducial area used in the present analysis ($25\times26\,\rm{cm}^2$) is $1.69\times10^4\,\rm{cm}^{-2}/\rm{fb}^{-1}$. The measured rate in the same area during the 2022 run amounts to $(2.07\pm0.01\,(stat)\pm0.10\,(sys))\times10^4\,\rm{cm}^{-2}/\rm{fb}^{-1}$, in good agreement with  predictions, thus validating the Monte Carlo simulation~\cite{simona}. The corresponding total number of muons integrated in $36.8\,$fb$^{-1}$ therefore amounts to $5.0\times10^8$, with $4.0\times10^8$ muons expected.

The inefficiency of the veto system planes is estimated from data by using good quality tracks reconstructed in the SciFi detector and validated with a track segment in the DS detector. The tracks are extrapolated to the veto detector fiducial volume. All tracks are identified as muons due to the large number of interaction lengths traversed; tracks entering the detector from the downstream end are excluded by timing measurements. 
For the first period of data taking amounting to $23.1\,$fb$^{-1}$, the applied time alignment procedure was relatively rough, leading to some physical events being split into two different recorded events. If one of the two does not contain enough hits to pass the online noise filter, this results in an apparent inefficiency of the detector. Therefore, for this period the measured inefficiency of a single plane is $8 \times 10^{-4}$, dominated by this effect. 
The problem was fixed at the end of October 2022 and the single-plane inefficiency dropped to $4 \times 10^{-5}$~\cite{trufvseff}. 
With the same method we have also estimated the inefficiency of the coincidence of the two veto detector planes, amounting to 
$7.1 \times 10^{-4}$ in the first period ($23.1\,$fb$^{-1}$) and $2.5 \times 10^{-6}$ in the second period ($13.7\,$fb$^{-1}$). The measured inefficiency of the double layer does not scale as the square of the single plane. The apparent correlation between the inefficiency of the two veto detector planes may be due to tracking imperfections in the inefficiency measurement or residual effects of the noise filter, both of which are expected to improve in the future. The overall veto system inefficiency during the 2022 run therefore amounts to $4.5 \times 10^{-4}$.



The SciFi detector inefficiency is estimated with a similar method used for the veto detector, using reconstructed SciFi tracks confirmed with a DS track and hits in the veto system. The presence of all SciFi stations is not required in the reconstruction, therefore the inefficiency of the first or second SciFi stations can be extracted. The inefficiency found for each station is $1.1\times 10^{-4}$.
The combined inefficiency of the veto system and the two most upstream SciFi planes is therefore $5.3\times10^{-12}$, thus making the background induced by muons entering the fiducial volume negligible.

Neutral particles (mainly neutrons and \KL's) originating from primary muons interacting in rock and concrete in front of the detector can potentially mimic a neutrino interaction since they do not leave any incoming trace in the electronic detectors, and can create a shower in the target associated with a DS track produced by punch-through or decay-in-flight $\pi^{\pm}$ and $K^{\pm}$. Although they are mainly rejected due to accompanying charged particles originating from the primary muon interaction, they constitute the main background source for the neutrino search. 

\begin{figure}[b]
\centering
\includegraphics[width=0.4\textwidth]{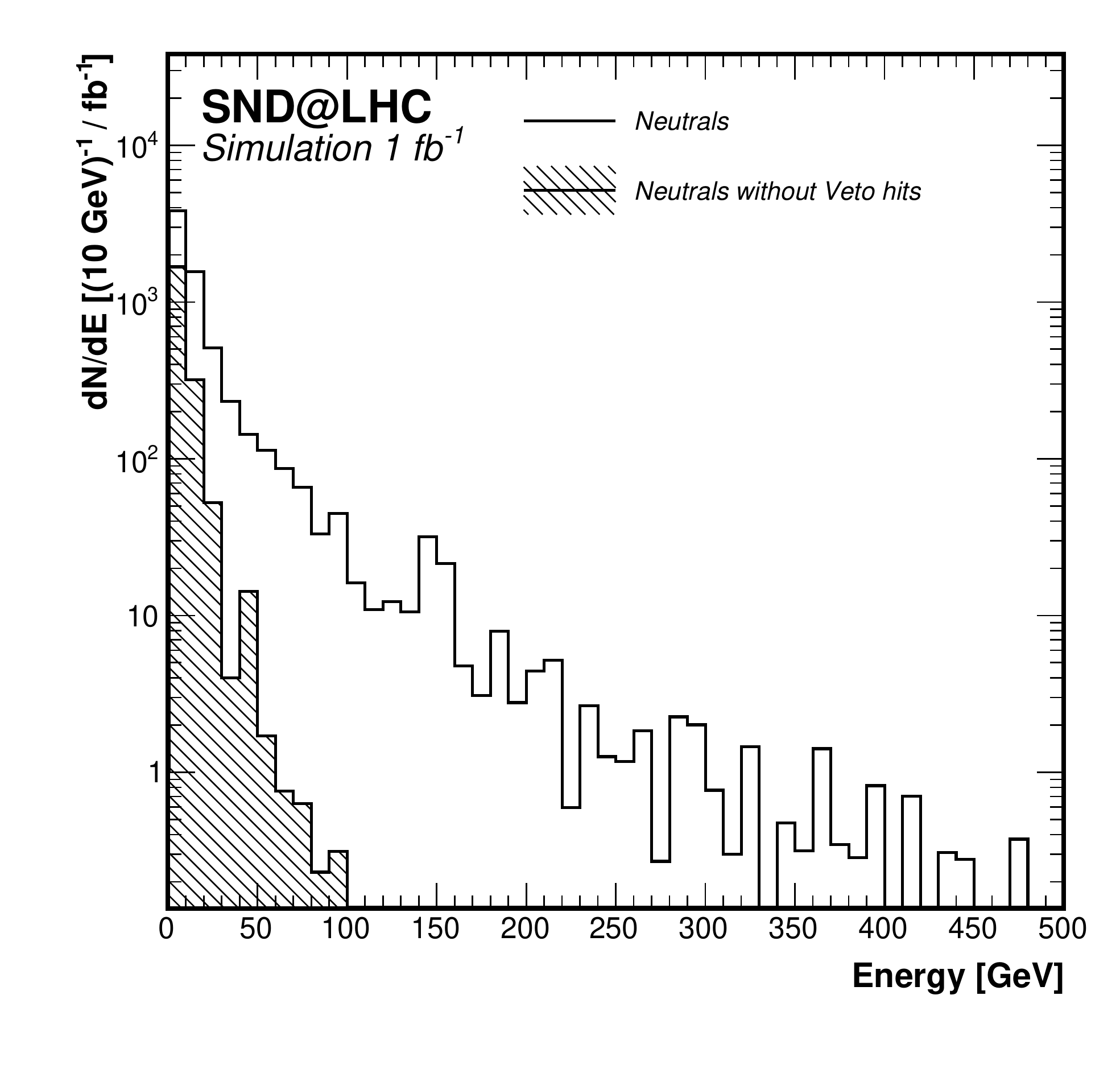}
\caption{Energy spectrum of neutral hadrons produced by muon interactions in the rock and concrete entering the \SND acceptance. The shaded area shows the spectrum after rejecting events with hits in the veto detector.}
\label{fig:neutral_energy_spectrum}
\end{figure}

\textsc{Pythia v6.4}~\cite{Pythia6} was used to simulate interactions of \mup and \mun on protons or neutrons at rest using the muon spectrum expected at the detector location. These events are placed along the muon flight direction according to the material density, and the secondary particles are transported by \textsc{Geant4} in the detector surroundings.
Neutral particles induced by muon DIS interact in the rock and concrete and only a small fraction of the particles leaves the tunnel wall and enters the detector. The energy spectrum of neutral hadrons entering the detector is shown in Figure~\ref{fig:neutral_energy_spectrum}, where the suppression achieved by rejecting events in which accompanying charged particles produce hits in the veto detector is also shown. 

To estimate the yield of neutral particles passing the event selection criteria, we simulate the highest energy neutral hadrons entering the target region in a given muon DIS interaction using \textsc{Geant4}~\cite{Geant4}. The events are simulated with energies within $[5,\,200]\,$GeV and uniformly distributed across the front face of the detector's target. As shown in Figure~\ref{fig:neutral_energy_spectrum}, the rate of neutral-hadron events with energies above 100 GeV is heavily suppressed by using the veto system to tag the accompanying charged particles (most often the scattered muon). Below 5 GeV the minimum ionizing particles resulting from the neutral hadron interactions do not have enough energy to produce a track exiting the downstream end of the detector.

As can be seen in Figure~\ref{fig:scifi_hits}, the lower energy of the neutral hadrons compared to the neutrino signal results in fewer hits in SciFi. We note that while this variable has not been used to reduce the neutral-hadron contamination in the present analysis, it is shown to be a powerful discriminant against this background.

The background yield after the selection amounts to $(7.6 \pm 3.1) \times 10^{-2}$ and is dominated by neutrons and $K^{0}_{L}$s. The systematic uncertainty of 40\% on the expected background yield is the combined effect of three sources. Since the neutral hadrons are produced in interactions of muons in the rock, we take as the uncertainty on the muon flux the 22\% difference between the simulated and measured flux of muons traversing detector. To estimate the impact of the hadron interaction model on the selection efficiency of these background events, we compare the results of simulations using two \textsc{Geant4} physics lists, \textsc{QSGP\_BERT\_HP\_PEN} and \textsc{FTFP\_BERT}, corresponding to two rather different hadron-nucleus interaction models, which differ by 22\%. Finally, the contribution to the systematic uncertainty due to the available statistics in the simulations is 25\%.



\vspace{0.3cm}

{\it Significance evaluation -} The significance of the observation of 8 candidates with an expected background yield of $7.6 \times 10^{-2}$ is quantified in terms of the exclusion of the null hypothesis, defined by setting the neutrino signal strength, $\mu$ to zero.
The one-sided profile likelihood ratio test $\lambda(\mu)$ was used as test statistic. The significance is evaluated by comparing $\lambda_{data}(\mu=0)$  with the sampling distribution of $\lambda(\mu=0)$. The likelihood, which includes a Gaussian factor to account for the background uncertainties, is
\begin{displaymath}
\mathscr{L} = \mathrm{Poisson}(n\,|\, \mu s + \beta)\, \mathrm{Gauss}(\beta\,|\,b,\sigma_b)
\end{displaymath}
where $n$ is the number of observed events, $s$ is the expected number of signal events and $\beta$ is the number of background events given by the Gaussian model, having a mean value $b$ and an uncertainty $\sigma_b$.
The implementation of the method based on RooStats\cite{rootstats} results in a $p$-value of $1.48\times10^{-12}$, corresponding to an exclusion of the background-only hypothesis at the level of $7.0$ standard deviations.

\label{sec:conclusions}

\vspace{0.3cm}

{\it Conclusions - } 
A search for high energy neutrinos originating from $pp$ collisions at $\sqrt{s} = 13.6\,$TeV is presented using data taken by the electronic detectors of \SND. We observe 8 candidate events consistent with \neum CC interactions. Our muon-induced and neutral-hadron backgrounds for the analysed data set amount to $(7.6 \pm 3.1) \times 10^{-2}$ events, which implies an excess of \neum CC signal events over the background-only hypothesis of seven standard deviations.

\vspace{0.3cm}

{\it Acknowledgments -} We express our gratitude to our colleagues in the CERN accelerator departments for the excellent performance of the LHC. We thank the technical and administrative staffs at CERN and at other SND@LHC institutes  for their contributions to the success of the SND@LHC effort. In addition, we acknowledge the support for the construction and operation of the SND@LHC detector provided by the following funding agencies:  CERN; the Bulgarian Ministry of Education and Science within the National
Roadmap for Research Infrastructures 2020–2027 (object CERN); the German Research Foundation (DFG, Deutsche Forschungsgemeinschaft); the Italian National Institute for Nuclear Physics (INFN); JSPS, MEXT, the Global COE program of Nagoya University, the Promotion
and Mutual Aid Corporation for Private Schools of Japan for Japan;
the National Research Foundation of Korea with grant numbers 2021R1A2C2011003, 2020R1A2C1099546, 2021R1F1A1061717, and 
\\2022R1A2C100505; Fundação para a Ciência e a Tecnologia, FCT (Portugal), CERN/FIS-INS/0028/2021, 
PRT/BD/153351/2021; the Swiss National Science Foundation (SNSF); TENMAK for Turkey (Grant No. \\2022TENMAK(CERN) A5.H3.F2-1).

M.~Climesu and R.~Wanke are funded by the Deutsche Forschungsgemeinschaft (DFG, German Research Foundation), project 496466340. We acknowledge the funding of individuals by Fundação para a Ciência e a Tecnologia, FCT (Portugal) with grant numbers
\\CEECIND/01334/2018, CEECINST/00032/2021 and PRT/BD/153351/2021. We thank Jakob Paul Schmidt and Maik Daniels for their help during the construction.

\onecolumngrid

\nocite{*}
\bibliography{apssamp}

\end{document}